\documentclass[]{iopart}
\usepackage{bm}
\usepackage{graphics}
\usepackage{graphicx}
\usepackage[dvips,usenames]{color}

\newcommand{\pd}[2]{{\frac{\partial #1}{\partial #2}}}

\newcommand{\be}{\begin{equation}}
\newcommand{\ee}[1]{\label{#1} \end{equation}}
\newcommand{\ba}{\begin{eqnarray}}
\newcommand{\ea}[1]{\label{#1} \end{eqnarray}}
\newcommand{\nl}{\nonumber \\}

\begin{document}
\article[Distributed mass QM]{Conference on Strange Quark Matter 2006}{
 Equation of state for distributed mass quark matter
}

\author{T S Bir\'o, P L\'evai, P V\'an and J Zim\'anyi}

\address{KFKI Research Institute for Particle and Nuclear Physics, H-1525 Budapest P.O.Box 49,
 Hungary}

\ead{tsbiro@sunserv.kfki.hu}

\begin{abstract}
We investigate how the QCD equation of state can be reconstructed by a continous
mass distribution of non-interacting ideal components. We find that adjusting the
mass scale as a function of the temperature leads to results which are conform to
the quasiparticle model, but a temperature independent distribution also may fit
lattice simulation results. We interpret this as a support for the quark
coalescence approach to quark matter hadronization.
\end{abstract}


\submitto{\JPG}


\section{Distributed mass quark matter}


The quark gluon plasma, produced in the Big Bang or in high energy relativistic heavy ion collisions,
hadronizes. We may detect and study this particular matter by observing
signs of its collective behavior, in local equilibrium obtaining its
equation of state. Several statistical and hydrodynamical models
have been considered in the soft QCD sector to describe hadron spectra.

We have developed a massive quark matter coalescence picture\cite{ALCOR} which determines hadron
yields and transverse spectra according to branching ratios between concurrent 
hadronization channels. Partonic level models of heavy ion reactions
also utilized the quark coalescence picture recently \cite{Mueller,CMKo,Denes}.
The seeming entropy reduction problem by coalescence with an associated reduction
(confinement) of color degrees of freedom can be resolved by assuming sufficiently
massive partons around the hadronization temperature in the precursor matter.
The necessary mass scale for quarks is about $300-350$ MeV and even higher (about $700$ MeV)
for gluons. In \cite{ALCOR} the explicit gluonic degrees of freedom were neglected.

In order to compose low hadron masses from massive constituent quarks 
we have introduced distributed mass partons into our hadronization model \cite{JPG2005}.
The low mass can be obtained from the convolution of distributed masses.

This approach led us to the investigation of an ideal but distributed mass parton gas. 
In particular we study i) the equation of state (EoS) for a continuous mixture,
ii) the consistency of the quasi-particle picture,
iii) the fit of different mass spectra to lattice eos data,
and iv) we collect arguments in favor of a mass gap.
We make a few interesting comparisons to non-ideal gas effects:
a fixed mass $M=gT$, scaling linearly with the temperature
in the quasi-particle picture, is known to lead to a reduced pressure compared to the ideal gas
(Stefan-Boltzmann) case even at infinite temperature $p/p_{SB} < 1$  
\cite{Peshier,BiShaTon,Tawflik}.
This  approach has been motivated by
partially resummed high-temperature pQCD calculations. 
A pressure reduction even at high temperatures 
also occurs in non-perturbative lattice QCD calculations. 

Slightly below $T_c$ the pressure does not vanish exactly. While earlier it was attributed to
the finiteness of the modeled lattice, modern scaling techniques along the line of constant
physics made us to believe that the nonzero pressure here is a real effect. In fact
massive resonance gas eos fits quite nicely this emerging part in the temperature range
$(0.8 - 1.1)T_c$ [Redlich,Tawflik].
We re-evaluate these phenomena in the light of the distributed parton mass model.

The particle spectrum is given as a convolution integral of the
spectral function $\rho(s)$  and a statistical
factor $f$. The former may be characterized
by dynamical factors, as e.g. a mass scale $M$, the latter by the
temperature $T$ and chemical potential $\mu$ which are characteristic
to the medium. 
A quasiparticle is described by a delta function with an arbitrary
dispersion relation $E_p$:
 $ \rho(E,\vec{p}) =  \frac{1}{2E_p} \delta\left(E-E_p\right).$
The distributed mass parton is equivalent to a continuous, finite-width
spectral function:
$  \rho(s) = \int\!dm \, w(m) \delta(m^2-s) = \frac{w(\sqrt{s})}{2\sqrt{s}}$.
An ansatz for $w(m)$ is equivalent to an ansatz for $\rho(s)$.
We note that field theoretical in-medium spectral functions break the Lorentz
covariance and show a separate dependence on the energy $E$ and momentum
$\vec{p}$. 

\section{Consistent equation of state with mass distribution}


Thermodynamical consistency of the quasiparticle picture imposes further
constraints on the mass distribution, $w(m)$. This can best be seen
when starting with a homogeneous equation of state, given as a continuous
sum of partial pressure contributions supported by a mean field part:
\be
  p(T,\mu) = \int\! dm \, w(m) \, p_m(T,\mu) - \Phi(T,\mu).
\ee{TOTAL-PRESS}
Here the partial contributions are given by the ideal gas formula at a fixed mass, 
\be
 p_m(T,\mu) = \int\! \frac{dE_k}{2\pi^2} \, \frac{k^3}{3} f\left(\frac{E_k-\mu}{T}\right)
\ee{PARTIAL-PRESS}
with $f(x)$ either the Fermi or the Bose (or 
approximately the Boltzmann-Gibbs) distribution and
with a particular, $m$-dependent dispersion relation for a free particle
$ k = \sqrt{E_k^2-m^2}$.
The entropy density and number density, as respective derivatives of the
pressure now contain extra terms due to the $T$- and $\mu$-dependence of the
mass distribution and of $\Phi$,
resulting in the following energy density, $e=Ts+\mu n-p$:
\be
\fl
  e =  \int\!dm\, w(m) \, e_m(T,\mu) + \int\!dm \, 
  	T\pd{w}{T} \, p_m + \mu\pd{w}{\mu} \,p_m
	 + \left(\Phi - T\pd{\Phi}{T}-\mu\pd{\Phi}{\mu} \right). 
\ee{ENERGY}
The quasiparticle consistency requires, that the total energy is also a sum
of the respective individual contributions plus the mean field contribution, $\Phi$. 
Therefore the mass distribution
$w(m)$ has to satisfy nontrivial constraints 
\be
 \int\!dm \, \pd{w}{T} \, p_m(T,\mu) = \pd{\Phi}{T}, \qquad
 \int\!dm \, \pd{w}{\mu} \, p_m(T,\mu) = \pd{\Phi}{\mu}.
\ee{CONSTRAINT}
In order to obtain $\Phi(T,\mu)$ the integrability condition,
$\partial_{T\mu}^2\Phi = \partial_{\mu T}^2\Phi$, has to be satisfied.
This, assuming that $w(m;T,\mu)$ is integrable, leads to
\be
  \int\!dm \, \pd{w}{T} \, n_m(T,\mu) \, = \, \int\!dm \, \pd{w}{\mu} \, s_m(T,\mu).
\ee{INTEGRABILITY}
The trivial way to satisfy this is to use a $T$- and $\mu$-independent mass
distribution, $w(m)$. In this case the mean field part, $\Phi$, is also
$T$ and $\mu$-independent and reduces to an old-fashioned bag constant.
The next step is the reconstruction of the equation of state. The pressure
(and energy density) modification $\Phi(T,\mu)$ is obtained from integrating
the constraint equations (\ref{CONSTRAINT}). 
The total pressure (eq.\ref{TOTAL-PRESS}) becomes
\be
 p(T,\mu) = \int_{T_c}^{T}dT \, \int w(m,T,\mu) s_m(T,\mu) dm - \Phi(T_c,\mu).
\ee{TOT-PRESS}
The mean field term, $\Phi$, cancels in the combination of $e+p$.
This fact can help one to guess an appropriate mass distribution
consistent with lattice eos data, as well as with a quasiparticle picture.

We consider from now on a particular class of mass distributions, which depend
on the thermodynamical environment parameters $T$ and $\mu$ only through a single
mass scale, $M(T,\mu)$:
\be
 w(m,T,\mu) = \frac{1}{M} \: f(\frac{m}{M}).
\ee{PARTICULAR-MASS-DISTRIBUTION}
The normalization integral for $w$ is inherited by the shape (form factor)
function $f(t)$:
\be
 \int w_m dm = \int f(t) dt = 1.
\ee{SPEC-MASS-INTEGRAL}
All medium dependencies are concentrated on $M(T,\mu)$, which should satisfy
the constraint stemming from the integrability condition eq.(\ref{INTEGRABILITY}).

\vspace{2mm}

Now we try to guess the proper mass distribution in order to arrive at a pressure
resembling lattice QCD simulation results.
A particular, analytically integrable ansatz for the mass distribution shape is given by
\be
  f(t) = A \exp\left(-at-b/t \right).
\ee{LEVAI-MASS}
Its normalization can be obtained from the general formula
\be
 \int_0^{\infty}dt \: t^{p-1} e^{-at-b/t} = 2 (b/a)^{p/2} K_p(2\sqrt{ab}).
\ee{KBES-EXP}
The pressure contribution normalized by the Stefan-Boltzmann pressure,
$p_{SB}=p_0$,
\be
 \sigma(g) = {\int w(m) \frac{p_m}{p_{SB}} dm} = \int_0^{\infty}dt \: f(t) \frac{g^2t^2}{2} K_2(gt), 
\ee{SIGMA}
with $g=M/T$, is given by
\be \fl
 \sigma(g) = g^2  \frac{\partial ^3}{\partial a^3} 
    \left[ K_2(\sqrt{b}\left(\sqrt{a+g}+\sqrt{a-g}\right))
	K_2(\sqrt{b}\left(\sqrt{a+g}-\sqrt{a-g}\right)) \right].
\ee{SIGMA-LEVAI}

\begin{figure}
\centerline{
\includegraphics[width=0.35\textwidth,angle=-90]{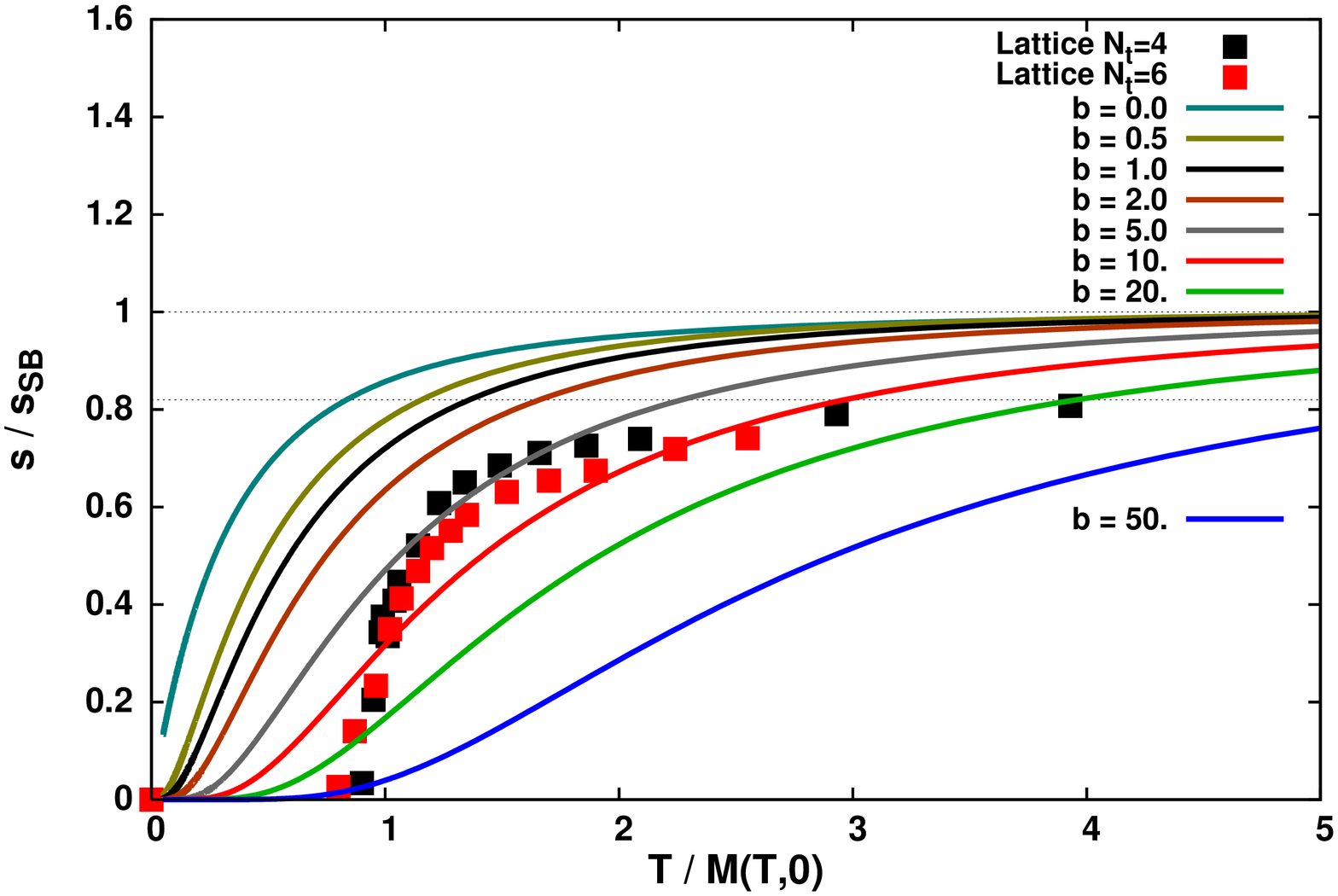}\includegraphics[width=0.35\textwidth,angle=-90]{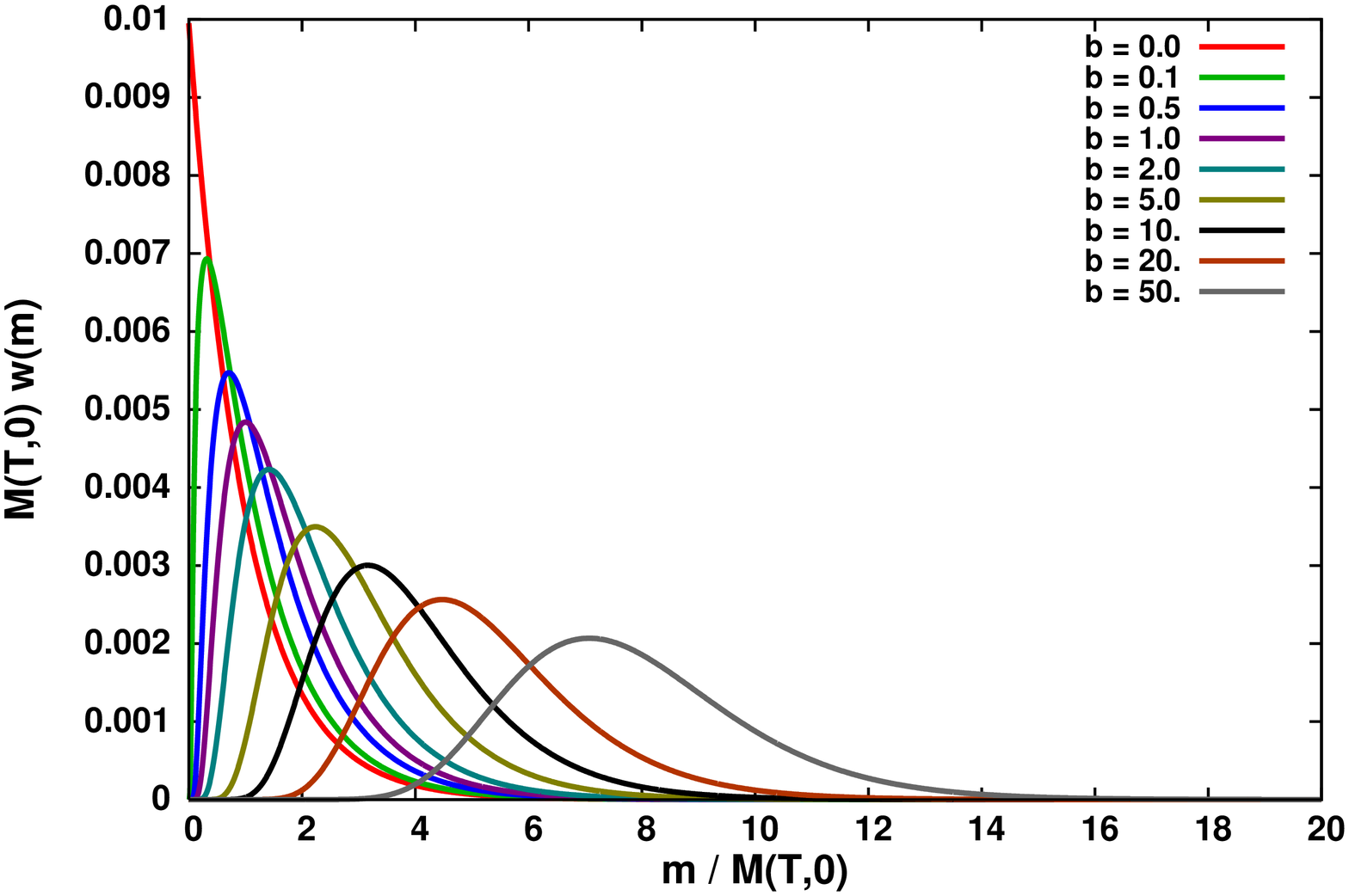}
}
\caption{ \label{FIG1:EoSab+MassDistab}
  (Color online)
  Normalized entropy -- temperature curves (left) 
  for different parameters of the $\exp(-am/T_c - bT_c/m)$
  mass distributions (right) compared with lattice QCD results\cite{AokiFodor2005} (boxes left).
}
\end{figure}

In \fref{FIG1:EoSab+MassDistab} we show the entropy density normalized by
the Stephan-Boltzmann value (belonging to the massless ideal gas), $s/s_{SB}=(e+p)/(e+p)_{SB}$ 
as function of the temperature (a) and the corresponding mass distributions (b).


\section{Fits to lattice QCD results}


\noindent
The lattice data of $p/p_{SB}$ can be related to $\sigma(g)$ by fitting the results.
Since due to eq.(\ref{SIGMA}) \hbox{$\sigma(0)=\int_0^{\infty} f(t)dt$,} the normalization
of the mass distribution requires $\sigma(0)=1$. In this case $p=p_{SB}$ would be
satisfied at infinite temperature ($g=T_c/T$).
As a matter of fact lattice data reach only $p/p_{SB}=0.8$ at $T \approx 4T_c$
instead. The quasiparticle model with a fixed mass which is growing proportional to
the temperature actually predicts such a deviation even in the infinite temperature
limit: in this case $\sigma(0)<1$. Therefore it is important to jugde without
prejudice whether lattice QCD data suggest a $\sigma(0)$ value lower than one or not.

In our understanding
a plot in terms of $g=T_c/T$ mediates a better picture of the exctrapolation to
the infinite temperature point at $g=0$. The lattice QCD eos data  do not contradict to
$\sigma(0)=1$. It is of course still imaginable that a bending down occurs
at high values of $T$ (low values of $g$) not simulated so far.
The tendency with growing lattice size (i.e. the comparison of $N_t=4$ and $N_t=6$ data),
however, seems to support $\sigma(0)=1$, a tautologic consequence of the distributed
mass model.
While this high-temperature behavior is well fitted by the pure
exponential $\sigma(g)$ function, the part below $T_c$ cannot be recovered this way.
A more steeply rising trial function is needed.

\begin{figure}
\centerline{
\includegraphics[width=0.40\textwidth,angle=-90]{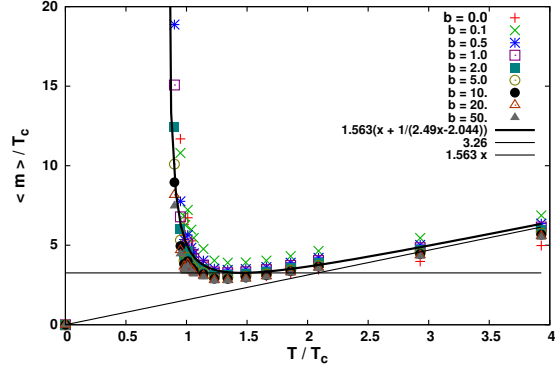}
}
\caption{ \label{FIG2:MexpT.ps}
 (Color online)
 The expectation value of the constituent mass as a function of the temperature as it follows
 from adjusting the mass distribution width parameter to lattice equation of state data.
}
\end{figure}

\Fref{FIG2:MexpT.ps} plots the expectation value of the mass as a function of
the temperature for the mass distribution \eref{LEVAI-MASS} obtained by adjusting
the pressure to the lattice result.


\begin{figure}
\centerline{
 \includegraphics[width=0.35\textwidth,angle=-90]{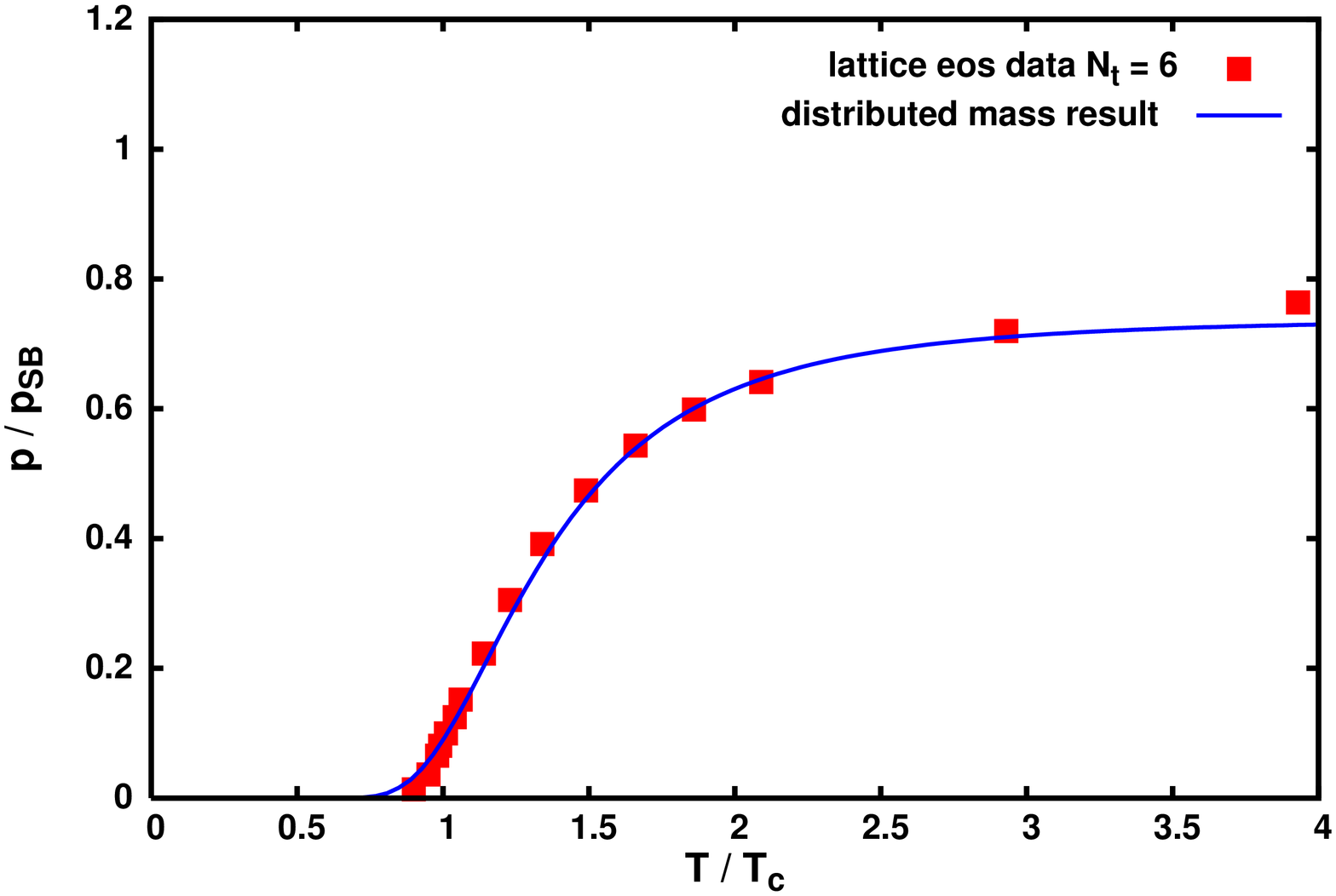}\includegraphics[width=0.35\textwidth,angle=-90]{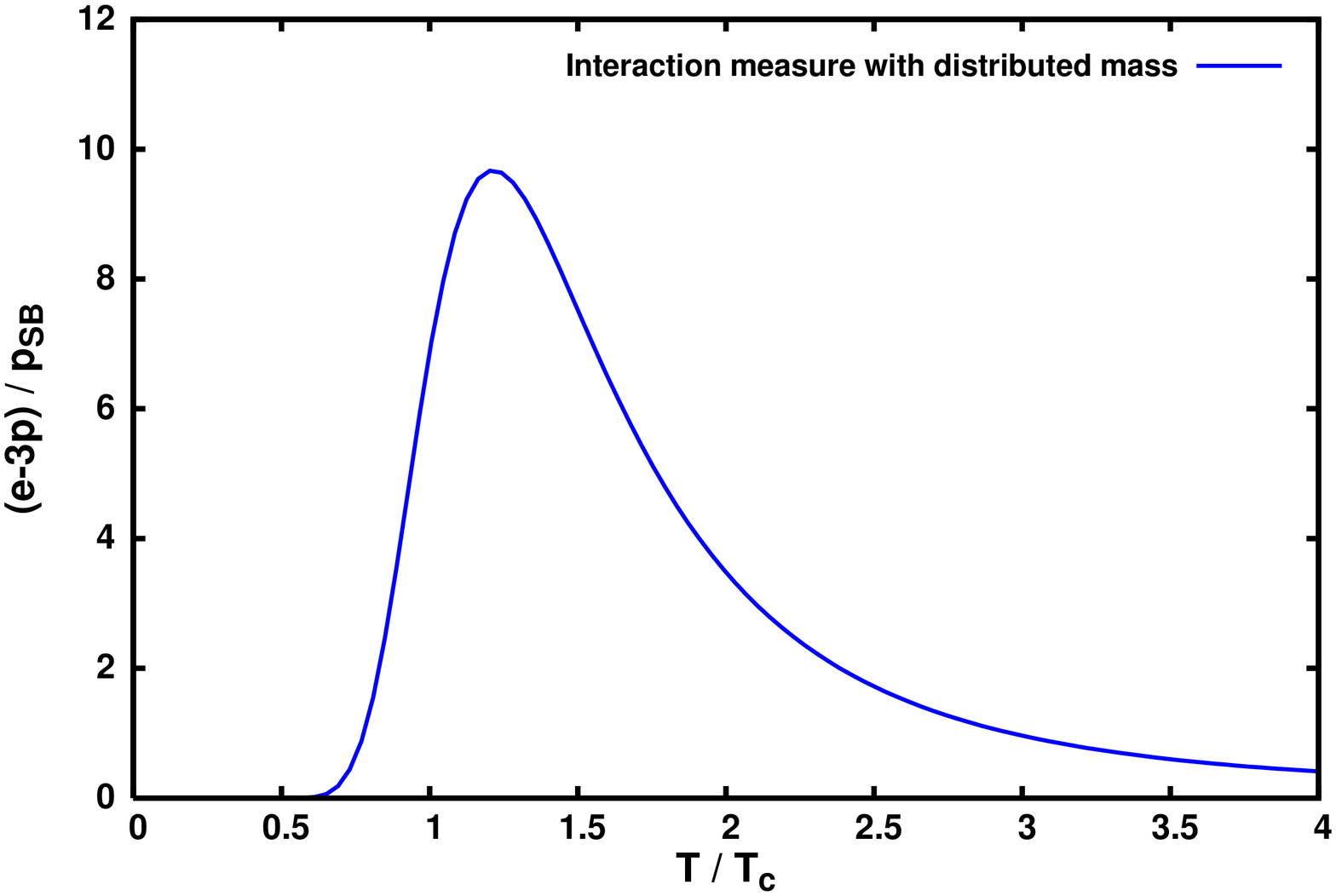}
}
\caption{ \label{FIG3:nyomas+em}
 (Color online)
 The normalized pressure (left) and the interaction measure
 $( \epsilon-3p)/T^4$ (right) obtained by using the mass distribution \eref{LEVAI-MASS}
 with adjusted mass sacle $M(T)$.
}
\end{figure}

\begin{figure}
\centerline{
 \includegraphics[width=0.44\textwidth,angle=-90]{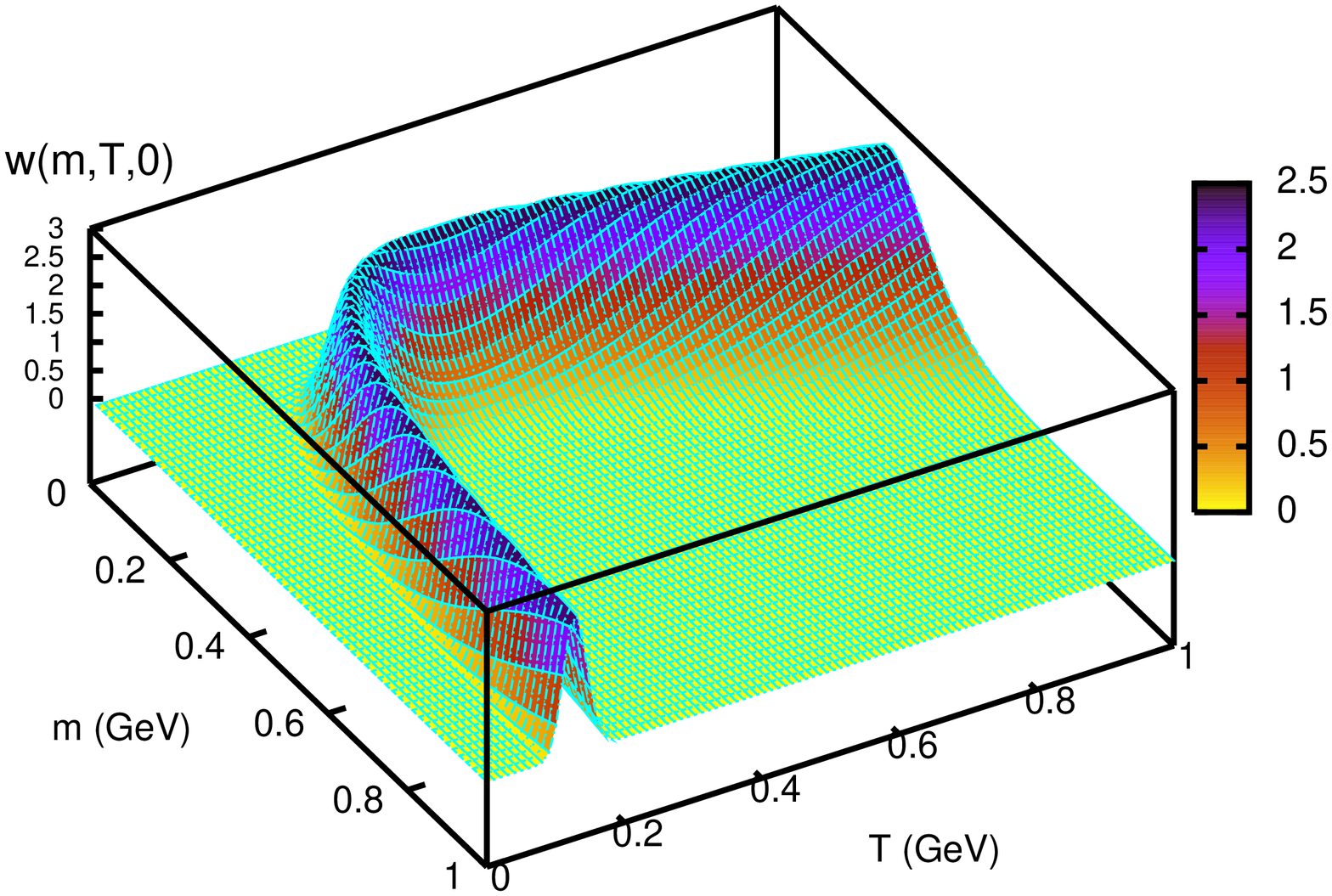}
}
\caption{ \label{FIG4:wmT}
 (Color online)
 The mass distribution for different temperatures, $w(m,T)=f(m/M(T))/M(T)$,
 with an adjuted mass scale to lattice QCD pressure results. It has allover
 a finite width, its local maximum follows the behavior of temperature dependent
 mass known from the quasiparticle model. $T_c=165$ MeV has been used.
}
\end{figure}

\Fref{FIG3:nyomas+em} presents the fitted pressure to lattice data and the interaction
measure, $(e-3p)/p_{SB}$.
The mass distribution \eref{LEVAI-MASS} as a function of the mass $m$ and the
temperature $T$ is plotted in \fref{FIG4:wmT}.

At the end of
this section we consider the mass distribution (\ref{LEVAI-MASS}) and investigate
the temperature and baryochemical potential dependence of its mean mass scale parameter,
$M(T,\mu)$.

The explicit showing of the other parameter(s), i.e. $b$, we shall suppress in the followings:
by plotting the expectation value of the mass instead of $M$ this is well founded.
The entropy density becomes in this special case
\be
 s = \pd{p}{T} = \frac{A}{\pi^2} T^3  
  \left(4\sigma\gamma - \frac{\mu}{T}\sigma\gamma' 
  +\left(- \frac{M}{T}\sigma' + \pd{M}{T}\right)\gamma\sigma' \right)
  - \pd{\Phi}{T}.
\ee{SPEC-ENTROPY-DENS}
The quasiparticle consistency is equivalent to the principle that all terms, which
were not there for a constant-mass, no-mean-field calculation, cancel:
\be
 \frac{A}{\pi^2} T^3 \pd{M}{T} \gamma \sigma' = \pd{\Phi}{T},
\ee{SPEC-T-CONSISTENCY}
leaving us with
\be
 s = \frac{A}{\pi^2} \left(4\sigma\gamma-\frac{\mu}{T}\sigma\gamma'
  - g\gamma\sigma' \right)
\ee{SPEC-CONSISTENT-ENTROPY-DENS}
Similarly the number density (here a conserved number density,
in QGP the one third of the baryon charge density, i.e. the quark minus
antiquark density) becomes
\be
 n = \pd{p}{\mu} = \frac{A}{\pi^2} T^3 \left(\sigma\gamma' + 
     \pd{M}{\mu}\gamma\sigma' \right) - \pd{\Phi}{\mu}.
\ee{SPEC-BARYON-THIRD-DENS}
The second quasiparticle consistency equation reads as
\be
 \frac{A}{\pi^2} T^3 \pd{M}{\mu} \gamma \sigma' = \pd{\Phi}{\mu},
\ee{SPEC-MU-CONSISTENCY}
leaving us with
\be
 n = \frac{A}{\pi^2} T^3 \sigma \gamma'
\ee{SPEC-CONSISTENT-DENS}
The energy density is combined to be
\be
 e = Ts + \mu n - p = \frac{A}{\pi^2} T^4 \left(3\sigma-g\sigma' \right)\gamma + \Phi.
\ee{SPEC-CONSISTENT-ENERGY-DENS}
It is noteworthy (although known for long) that the combination $e+p$ is independent
of the mean field part:
\be
 \frac{e+p}{T^4} = \frac{A}{\pi^2} \gamma \left(4\sigma-g\sigma' \right)
\ee{E-PLUS-P}
Now we investigate the integrability condition in this special case.
The $\mu$-derivative of the $T$-consistency equation (\ref{SPEC-T-CONSISTENCY})
(we shall call left hand side, LHS) has to be equal to the $T$-derivative
of the $\mu$-consistency eq. (\ref{SPEC-MU-CONSISTENCY}). Omitting two common
terms in both, namely $\gamma\sigma'\partial^2M/\partial T \partial\mu$ 
and $\pd{M}{T}\pd{M}{\mu}\gamma\sigma'' $, we are left with
\ba
 LHS = \pd{}{\mu} \pd{\Phi}{T} &=& \frac{A}{\pi^2} T^2
  \gamma'\sigma' \pd{M}{T}, \nl
 RHS = \pd{}{T} \pd{\Phi}{\mu} &=& \frac{A}{\pi^2} T^2
 \left({3}\gamma\sigma' - \frac{\mu}{T}\gamma'\sigma'-\frac{M}{T}\gamma\sigma'' \right)
  \pd{M}{\mu}.
\ea{LHS-RHS}
from the equality of the above expressions it follows
\be
 \sigma'\gamma'\pd{M}{T}
 +\left(g\gamma\sigma'' + \frac{\mu}{T} \gamma'\sigma'-3\gamma\sigma'\right) \pd{M}{\mu} = 0.
\ee{SPEC-INTEGRABILITY}
In this partial differential equation for $M(\mu,T)$ $\sigma$ depends only
on $g=M/T$ and $\gamma$ depends only on $\alpha=\mu/T$. The solution can be 
characterized by $M(T,\mu)$=constant lines on the $T-\mu$ plane. Such curves follow
an ordinary differential equation obtained from (\ref{SPEC-INTEGRABILITY}):
\be
 \frac{d\mu}{dT} = \frac{\gamma}{\gamma'} \left(g\frac{\sigma''}{\sigma'}-3 \right)
 +\frac{\mu}{T}.
\ee{SPEC-CURVE}
The solution has the form $\mu=T\alpha(g)$ whence we obtain
\be
 \frac{1}{\gamma} \frac{d\gamma}{d\alpha} \left(\frac{d\mu}{dT}-\frac{\mu}{T} \right)
 = g \frac{\sigma''}{\sigma'} - 3.
\ee{THAT-SPEC-CURVE}
The expression in the bracket above reduces to
\be
 \frac{d\mu}{dT} - \frac{\mu}{T} = T \frac{d\alpha}{dT} = - g \frac{d\alpha}{dg}
\ee{THAT-BRACKET}
while $gT=M$ is constant. Therefore we obtain the following differential equation:
\be
 - \frac{g}{\gamma} \frac{d\gamma}{d\alpha} \frac{d\alpha}{dg} = g \frac{\sigma''}{\sigma} - 3,
\ee{THAT-DIFF-EQ}
which becomes an integrable problem for $\gamma(g)$:
\be
 - \frac{d}{dg} \ln \gamma = \frac{\sigma''}{\sigma'} - \frac{3}{g}.
\ee{THAT-INTEGRABLE}
Its solution is given by
\be
 \gamma = K g^3 / \sigma'(g).
\ee{THAT-SOLUTION}
This, replacing the definition of $\gamma$ in terms of $\mu/T$, 
and $g=M/T$ with constant $M=m_0$ one obtains a definite integral curve starting
at $T=T_0$ for $\mu=0$. That means that the constant $K$ is related to $T_0$.
By inverting the $\cosh$ function the following explicit solution emerges for
the $M=m_0$ constant lines:
\be
 \mu = T \ln\left(z(T)+\sqrt{z^2(T)-1}\right)
\ee{CURVE}
with
\be
 z(T) = \left(1+\frac{\gamma_G}{2\gamma_Q} \right) 
   \frac{T_0^3\sigma'(m_0/T_0)}{T^3\sigma'(m_0/T)} - \frac{\gamma_G}{2\gamma_Q}.
\ee{ZETOR}
The temperature dependence of the mass scale at $\mu=0$, $M(T,0)=m_0$, 
has to be inverted in order to obtain $T_0(m_0)$. This positions the crossings of $M=m_0$ constant
characteristics with the temperature axis. We note that in every model with baryon - antibaryon
symmetry these lines have zero $\mu$-derivatives at $\mu=0$.

\begin{figure}
\centerline{
 \includegraphics[width=0.50\textwidth,angle=0]{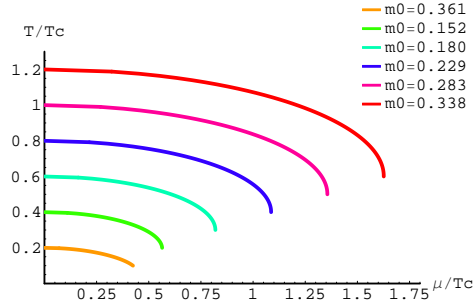}
}
\caption{ \label{FIG5:vanp}
 Continuation of the medium dependence of the single mass scale to finite
 chemical potential, $M(T,\mu)$, using a 12:16 light quark -- gluon mixture with mass
 distribution. It have been obtained by solving the integrability condition 
 \eref{SPEC-INTEGRABILITY}
 starting from the temperature dependent scale $M(T)$ fitted to lattice QCD results
 at zero chemical potential. The $m_0=M(T,0)$ values are given as $m_0=0.341, 0.129,
 0.146, 0.184, 0.226$ and $0.270$ GeV from above. 
}
\end{figure}

\section{Arguments for a mass gap}


The $\sigma(g)$ function in the Boltzmann approximation is given by \eref{SIGMA}.
This is a so called Meijer K-transform (a generalized Laplace transform)
which  can be inverted by
\be
 f(t) \: = \: \frac{2}{i\pi } \int_{c-i\infty}^{c+i\infty} \sigma(g) \: \frac{I_2(gt)}{gt} \, dg.
\ee{INV-MEIJER}
This presents a peculiar problem, whether
there exists a unique mass distribution $f(t)$, with the mass scale parameter kept
temperature and chemical potential independent, 
to any $\sigma(g)$ function extracted from an equation of state (e.g. from lattice QCD calculations). 
The shape of such a mass distribution is not arbitrary. 
We shall  explore this possibility in a future work.

Utilizing the parametric integral \eref{SIGMA} one easily derives a useful
relation between moments of the mass distribution, $f(t)$ and the eos fit.
We obtain
\be
 M_n = \langle t^{-n} \rangle_f \: = \: \frac{1}{\sqrt{\pi}} \, 
    \frac{\Gamma\left(\frac{n+1}{2}\right)}{\Gamma\left(\frac{n}{2}+2\right)\Gamma(n)} 
 \langle g^{n-1} \rangle_{\sigma} \:
  = \: c_n \langle g^{n-1} \rangle_{\sigma}.
\ee{MASS-EOS-MOMENTS}
A roughly approximate, qualitatively correct fit to the lattice eos data is represented
by the straight line, $ \sigma(g) = 1 - g/g_c$, with  $g_c\approx 1.15$. 
All the $n > 0$ moments of this expression are finite, meaning
that the inverse mass moments are also finite:
\be
 \int_0^{\infty} f(t) \frac{dt}{t^n} = M_n = \frac{c_ng_c^n}{n(n+1)}.
\ee{FINITE-MOMENTS}
This is possible only if the mass distribution has a finite mass gap, or it 
approaches zero more than any polynomial in the inverse mass $1/t$. Both possibilities
represent an interesting spin off of the eos studies.
The sizeable reduction of the pressure
at low temperature (large $g$), which causes the $n>0$ moments of $\sigma(g)$ be finite in general,
is related to confinement. This requires finite moments of the inverse mass with the
mass distribution, which is possible only with a corresponding suppression of the low mass
part, alike $\exp(-M/m)$, or with a mass gap.


\ack{
This work was supported by the Hungarian National Research Fund OTKA (T48489, T49466).
We are indebted for useful discussions to Prof. B. M\"uller (Duke University, NC USA),
Dr. A. Jakov\'ac (Technical University Budapest), Prof. A. Patk\'os and
Dr. Z. Sz\'ep (E\"otv\"os University Budapest).
}


\end{document}